\documentclass[fleqn,usenatbib]{mnras}

\usepackage{newtxtext,newtxmath}

\usepackage[T1]{fontenc}
\usepackage{listings}
\usepackage{xcolor}
\usepackage{lipsum}
\DeclareRobustCommand{\VAN}[3]{#2}
\let\VANthebibliography\thebibliography
\def\thebibliography{\DeclareRobustCommand{\VAN}[3]{##3}\VANthebibliography}

\usepackage{graphicx}	
\usepackage{amsmath}	

\usepackage{amssymb}	
\usepackage{color}
\usepackage{ulem}

\title[HCGrid for Gridding]{HCGrid: A Convolution-based Gridding Framework for Radio Astronomy in Hybrid Computing Environments}

\author[Hao Wang et al.]{
Hao Wang,$^{1}$
Ce Yu,$^{1}$
Bo Zhang,$^{2,3}$\thanks{E-mail: zhangbo@nao.cas.cn}
Jian Xiao$^{1}$\thanks{E-mail: xiaojian@tju.edu.cn}
and Qi Luo$^{1}$
\\
$^{1}$College of Intelligence and computing, Tianjin University, No.135 Yaguan Rood, Haihe Education Park, Tianjin, 300350, China\\
$^{2}$National Astronomical Observatories, Chinese Academy of Sciences, No.20 Datun Road, Chaoyang District, Beijing, 100012, China\\
$^{3}$CAS Key Laboratory of FAST, National Astronomical Observatories, Chinese Academy of Sciences
}

\date{Accepted XXX. Received YYY; in original form ZZZ}

\pubyear{2020}

\begin{document}
\label{firstpage}
\pagerange{\pageref{firstpage}--\pageref{lastpage}}
\maketitle
\newcommand\blfootnote[1]{%
\begingroup
\renewcommand\thefootnote{}\footnote{#1}%
\addtocounter{footnote}{-1}%
\endgroup
}
\begin{abstract}
 Gridding operation, which is to map non-uniform data samples onto a uniformly distributed grid, is one of the key steps in radio astronomical data reduction process.
 One of the main bottlenecks of gridding is the poor computing performance, and a typical solution for such performance issue is the implementation of multi-core CPU platforms. Although such a method could usually achieve good results, in many cases, the performance of gridding is still restricted to an extent due to the limitations of CPU, since the main workload of gridding is a combination of a large number of single instruction, multi-data-stream operations, which is more suitable for GPU, rather than CPU implementations.
 To meet the challenge of massive data gridding for the modern large single-dish radio telescopes, e.g., the Five-hundred-meter Aperture Spherical radio Telescope (FAST), inspired by existing multi-core CPU gridding algorithms such as Cygrid, here we present an easy-to-install, high-performance, and open-source convolutional gridding framework\textbf{, HCGrid,} in CPU-GPU heterogeneous platforms. It optimises data search by employing multi-threading on CPU, and accelerates the convolution process by utilising massive parallelisation of GPU. In order to make HCGrid a more adaptive solution, we also propose the strategies of thread organisation and coarsening, as well as optimal parameter settings under various GPU architectures. A thorough analysis of computing time and performance gain with several GPU parallel optimisation strategies show that it can lead to excellent performance in hybrid computing environments.

\end{abstract}




\section{Introduction}

In radio astronomy, gridding operation is a key step towards generating 2D sky maps or 3D cubes with evenly-spaced grids for scientific research or data release \citep{winkel2016cygrid}\blfootnote{\textit{This is a pre-copyedited, author-produced PDF of an article accepted for publication in} Monthly Notices of the Royal Astronomical Society \textit{following peer review. The version of record} Hao Wang, Ce Yu, Bo Zhang, Jian Xiao and Qi Luo, HCGrid: A Convolution-based Gridding Framework for RadioAstronomy in Hybrid Computing Environments, Monthly Notices of the Royal Astronomical Society, \textit{is available online at} \url{https://doi.org/10.1093/mnras/staa3800}}. However, the original observed data are usually not so uniformly distributed. Taking the planned Commensal Radio Astronomy FAST Survey (CRAFTS, see \citealt{Li2018}) of the Five-hundred-meter Aperture Spherical Radio Telescope (FAST) \citep{nan2006five,nan2011five} as an example. During this survey, the FAST telescope will be operated under the drift-scan mode, that is, the telescope will remain fixed along local meridian, with different patches of the sky entering the field of view due to rotation of the Earth. With a beam size of $\sim 2'.9$ at $\sim 1.42$ GHz, and rotation angle of $\sim 23.4$ degrees, the 19-beam receiver of FAST \citep{smith2017performance} can achieve a beam spacing of $\sim 1' 09''$ along declination, with super Nyquist coverage guaranteed. While on the right ascension direction, typically the telescope's spectral line backends can record data once every second \citep{Li2018}, with a maximum sampling rate of 10 times per second, thus providing a much denser coverage. Hence, gridding is a necessity to resample such observations uniformly, for the convenience of data analysis, visualisation or storage \citep[e.g.][]{lena2012observational}.

However, gridding is a computing and I/O intensive job \citep{giovanelli2005arecibo}, comprising one of the most time-consuming steps in the complete data reduction process. And for the latest instruments with high data production rates such as FAST, the speed of gridding operation should be of great concern. Our experiments have shown that with a 16-core CPU platform, it would take $\sim 21$ CPU hours to perform gridding on 1 TB raw data using the multi-thread method.
However, the CRAFTS project is expected to generate as many as $10\sim20$ PB-sized data every year \citep{Li2018}. That is, hundreds of CPU-hours are required to grid the raw data generated by FAST in real-time with the existing methods, which could greatly increase the cost of computation for this survey.

In the past decade, various studies on optimisation of gridding performance have been carried out. For example, \cite{winkel2016cygrid} have presented a versatile gridding module (\textbf{Cygrid}: A fast Cython-powered convolution-based gridding module for Python) for radio astronomical data reduction, featuring a single-level lookup table based on the C++ standard template library (STL) for gridding parallelisation, as well as the HEALPix-based spherical tessellation for fast neighbor searching. And the performance of the Cygrid code can be significantly improved when using multiple CPU cores.
However, utilising the STL vector, which features dynamical arrangements of storage allocations with a relatively small initial storage volume, to construct a lookup table for input raw data is not always practical. For example, if the raw samples are unevenly distributed, with more data points in the central regions and less on all sides, the required storage needs to be expanded in the middle part of each HEALPix-tessellated ring frequently. In this case, the STL vectors have to make massive data copies of the data frequently, which would consume broad bandwidth, thus resulting in degradation of gridding-performance.

In addition, the main workload of gridding can be considered as a combination of a large number of single-instruction, multi-data stream operations. For such tasks, the high parallelisation level, as well as throughput of GPUs brings a possibility for better performance. Thus, \cite{romein2012efficient} has developed a work-distribution scheme for radio data gridding with GPUs, which could significantly reduce the memory access time of computing device, and map observed samples onto a grid with high efficiency. And \cite{merry2016faster} has further optimised this algorithm with thread coarsening strategy, making one thread handle multiple samples simultaneously. However, the method proposed by \cite{romein2012efficient} and \cite{merry2016faster} has been deeply customised for radio telescope arrays, strongly relying on the spatial coherence of the interference matrix, thus cannot be easily adapted for single-dish telescopes.
   
Accordingly, in order to meet the requirements of large data volume from the latest single-dish radio telescopes, and to overcome the shortcomings of GPU-based gridding methods mainly designed for radio telescope array, inspired by \cite{winkel2016cygrid}, and based upon our previous work \citep{luo2018hygrid}, here we present a convolutional gridding framework with hybrid computing environments, \textbf{HCGrid}, for radio astronomy, which can greatly improve the performance of the gridding process. Key features of our framework can be summarised as follows:
   \begin{enumerate}
       \item We have designed a gridding module with high performance for large single-dish radio telescopes.
       \item  We partitioned the sample space based on HEALPix (Hierarchical Equal Area Latitude Pixelation, \citealt{gorski2005healpix}). To effectively utilise the storage resources, and to accelerate the searching process of effective contributing data points, we adopted the two-level lookup table scheme proposed by our previous work, which can further improve the degree of parallelism of gridding.
       \item We have proposed our scheme for thread organisation, thread coarsening, as well as data layout for further optimisation of the performance gain.
       \item We have conducted comparative experiments and comprehensive analysis with 3 types of mainstream GPU architectures, Kepler, Turing, and Volta, resulting in a detailed and easy-to-use performance optimisation guide for various scenarios of applications.
   \end{enumerate}
  
The rest of this paper is organised as follows. Section 2 presents the details of convolution-based gridding algorithms in radio astronomy and the corresponding parallelisation strategy. Section 3 focuses on the design of our framework. And the results of the related experiments are described in Section 4. Our conclusion is drawn in Section 5 with further discussions.

\section{Gridding algorithms in astronomy}

\subsection{Convolution-based gridding algorithm}

  The convolution-based algorithm is among the most common choice for gridding of radio astronomical data. For example, \cite{kalberla2005leiden,kalberla2010gass} have applied this technique to produce data cubes and HI column density maps for the Leiden/Argentine/Bonn Survey, while \cite{winkel2016effelsberg} have adopted a similar approach in the Effelsberg–Bonn HI Survey. The first step of convolutional gridding is to design a target grid, and then the resampled value for each grid cell is calculated.

The algorithm we adopted is mainly from \cite{winkel2016cygrid}. The output value for each targeted grid cell equals the weighted sum of all neighboring samples. Let $ \mathbb{S}=\{s_1,s_2,\cdots,s_N\}$ denote $N$ non-uniformly spaced samples distributed across the RA-Dec plane. For $n$th sample $s_n\in\mathbb{S}$, its equatorial coordinates should be expressed as $(\alpha_n,\delta_n)$ ($\alpha_n$ means the right ascension, $\delta_n$ the declination), with a sampled value of $V[s_n]$. For our target grid, suppose the RA-Dec plane is divided into a regular grid with $I \times J$ cells $\mathbb{G}=\{g_{1,1},g_{1,2},\cdots,g_{I,J}\}$. For any cell $ g_{i,j}\in\mathbb{G} $ with central coordinates $(\alpha_{i,j}, \delta_{i,j})$, its resampled value $V[g_{i,j}]$ is equivalent to the weighted sum of original data $\mathbb{S}$ related to $ g_{i,j} $ \citep{winkel2016cygrid}
     \begin{equation}
      V[g_{i,j}] = {\frac{1}{W_{i,j}}}
         \sum_{n}V[s_n]\omega(\alpha_{i,j},\delta_{i,j};\alpha_n,\delta_n)
         \label{convolution}
     \end{equation}
\noindent
Here $s_n$ represents any original input sample with a weighted contribution to $ g_{i,j} $; $\omega(\alpha_{i,j},\delta_{i,j};\alpha_n,\delta_n)$ is a convolution kernel (weighting function) depending on positions of the target cell and original data points, usually related to distances between input and output coordinates; and $W_{i,j}= \sum_{n}\omega(\alpha_{i,j},\delta_{i,j};\alpha_n,\delta_n)$ is the normalisation coefficient \citep{winkel2016cygrid}. Since each input sample can influence different output cells with different degrees, with the introduction of $\omega_{i,j}$, the resampled output values can be correctly calculated.
  
Generally speaking, the resolution limit of the target (output) grid $\sigma_{grid} $ (that is, the distance between any two adjacent cells) is determined by the instrument resolution $\sigma_{data} $) and the size of the convolution kernel function $\sigma_{kernel}$ as $\sigma_{grid} = \sqrt{\sigma_{data}^2 + \sigma_{kernal}^2}$ , with $\sigma_{kernel}\sim \frac{1}{2}\sigma_{data}$ as a common choice \cite{winkel2016cygrid}. Besides, since by adopting the coordinates of FITS \citep{wells1979fits} world coordinate system (WCS, see \citealt{calabretta2002representations}, as well as \citealt{mink2006wcstools}) in calculations, the gridding process can directly perform convolutions in WCS space, which has proved to be more convenient for astronomical users, we mainly focus on astronomy-specific cases of WCS-targeted gridding in the following discussions.

\subsection{Gridding parallelisations}
In order to speed up the computing process, the gridding algorithm is usually parallelised with multithreadings using multiple CPUs, or CPU-GPU heterogeneous computing environments. Two types of parallelisation strategies exist, scattering and lookup table-based gathering strategies.

The scattering strategy calculates the contribution of each input data point to all target cells within the influence of the gridding kernel, thus facing the risk of ``writing competition'', that is, conflict when calculating the same target cell value with different set of input data. Although efficient scheduling algorithms can be invoked to avoid the risk of such competitions \citep{mccool2012structured}, the algorithms themselves often show difficulties for GPU implementations, thus rendering the scattering method less effective when deploying under GPU environments \citep{van2009evaluating}. On the other hand, although atomic operations, which cannot be interrupted by thread scheduling, can also be utilised to avoid writing competitions, its could also make parallel-scatterings degenerate into serialised algorithms in the presence of densely sampled raw data or frequent competitions \citep{schweizer2015evaluating}, thus affecting the speed of gridding operations.

\begin{figure}
   \centering
   \includegraphics[width=5cm]{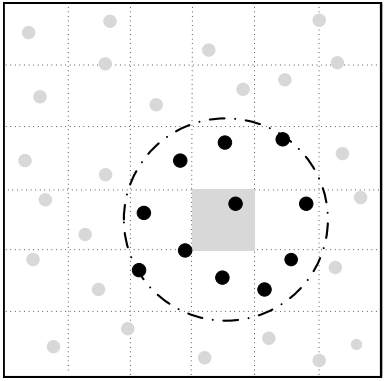}
      \caption{Gathering method: one approach of implementing convolution-based gridding.
                Each output data point (as shown with light-colored grid) collects its resampled value from all neighboring input data (darker dots) lying within a kernel (dotted circle).
                }
         \label{gather}
   \end{figure}

While as shown in Fig.~\ref{gather}, the gathering method identifies all adjacent input points within the range of influence of the gridding kernel for each target cell, and performs convolution with such contributors to calculate the final results. Although the writing competition issue no longer exists in this case, it is nearly impossible to locate all the input points within a certain kernel radius directly for any designated cell, due to the non-uniformity of the input samples. Thus, massive searching operations are required. To avoid unnecessary searching operations, and to improve the capability of the entire gridding process, the gather strategy usually performs pre-ordering the input data in advance, partitioning the sampled data according to their coordinate information, with the establishment of block number - sampled point lookup table (one-to-many mapping). After that, blocks falling into the convolution kernels of each grid cell can be quickly determined and accessed (see \cite{luo2018hygrid} and references herein). HEALPix \citep{gorski2005healpix} is a tessellation scheme for the spherical surface. Its properties make it a useful tool for the constructions of lookup-table schemes. Based on HEALPix, the gathering strategy can take more advantages of GPU, compared with the scattering one. Therefore, we adopt the gathering method in our gridding framework for CPU-GPU hybrid platforms. In fact, the main reference work of this paper, Cygrid \citep{winkel2016cygrid}, is also based on the gathering method.

%
\section{Details of the gridding framework implementation}

\subsection{Overview of HCGrid}
\begin{figure}
   \centering
   \includegraphics[width=8.5cm]{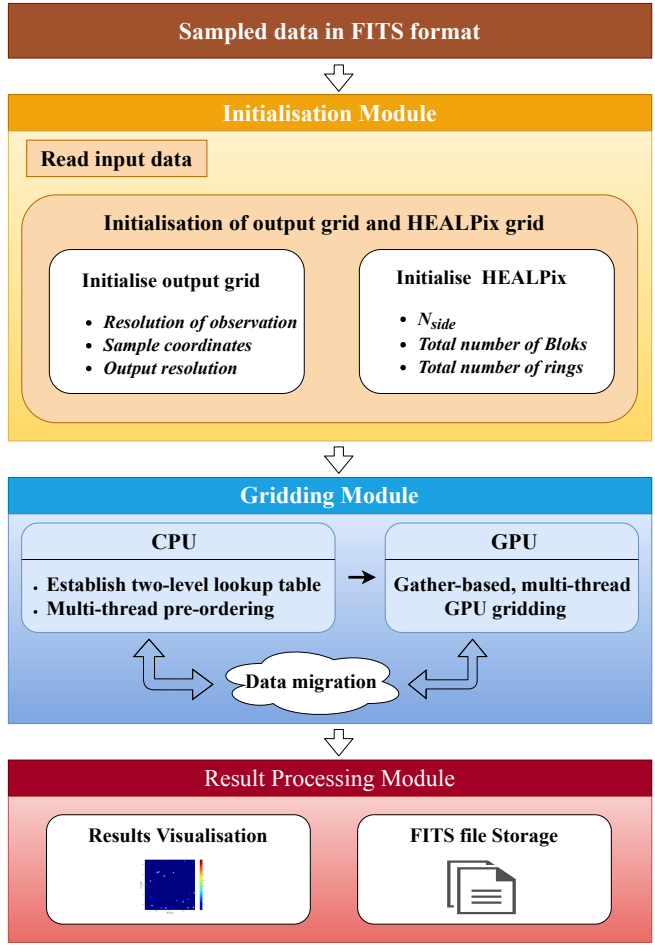}
      \caption{The architecture of HCGrid. HCGrid consists of three modules, including initialisation, which is mainly used for initialisation of parameters involved in the calculation process; gridding, which consists of the core of HCGrid; and finally, result processing for gridding output visualisation and data storage.}
         \label{Module}
   \end{figure}

  Fig.~\ref{Module} shows the basic architecture of HCGrid, which consists of three modules:
 \begin{enumerate}
     \item \textbf{Initialisation module} for raw data initialisation. This module imports FITS-format input data files, extracting related parameters (such as $ \sigma_{data} $, sample coordinates, as well as output resolution) from therein, and initialises the output grid and HEALPix with extracted parameters.
     \item \textbf{Gridding module}, which is the core module of the HCGrid framework. This module performs data pre-ordering on CPU, and gridding on GPU, along with data migration between CPU and GPU.
     \item \textbf{Result-processing module}, which visualise the gridding results, as well as exporting the final products as FITS files.
 \end{enumerate}
The gridding module as our main topic of interest for this work, we will mainly discuss the gridding module in the following sections. The source code of HCGrid can be accessed openly via GitHub\footnote{\url{https://github.com/HWang-Summit/HCGrid}}, and all suggestions and user's feedback are welcome.
 
\subsection{HEALPix-based partition on sampling space}
As mentioned in Section 2.2, the aim of input-data pre-ordering is to improve the searching efficiency on the contributing data points, as well as the performance of the gridding process, by means of rearrangements of the original samples. Here we adopt HEALPix for sampling space partition.

HEALPix \citep{gorski2005healpix} is a software package for hierarchical equal-area isolatitude pixelation on spherical surfaces, thus making fast, accurate statistical or astrophysical analysis of massive all-sky data sets possible. With the help of HEALPix-based raw-data space partition, one can determine pixel and ring indices of  raw data within a certain area, thus reducing the workload of searching operation, and enabling more reasonable distributions of raw data covering a spherical surface.
HEALPix provides two possible implementations of pixel indexations, the ring scheme and the nested scheme, arranged on isolatitude rings, or in a nested tree fashion, respectively. Both schemes can map input samples to a one-dimensional numbered sequence. Since with the ring scheme, the pixel indices increase strictly linearly along the latitude ring, which makes it easier to establish a lookup table based on positions of each pixel, here we choose to utilise such a scheme as basis for data pre-ordering and lookup-table construct.
  
The HEALPix software library supports coordinate conversions based on the WCS standard (\citealt{calabretta2002representations} and \citealt{mink2006wcstools}). Through related APIs of this library, one can make a series of pixel manipulations. For example, given a partition level "$N_{side}$",  we can obtain the corresponding ring index of a certain pixel, the world coordinate of the pixel center, the index number of the pixel, the starting pixel index of the ring, and so on. Thus, the indexing and searching-related operations can be implemented with these HEALPix APIs.

   \begin{figure}
   \centering
   \includegraphics[width=8.5cm]{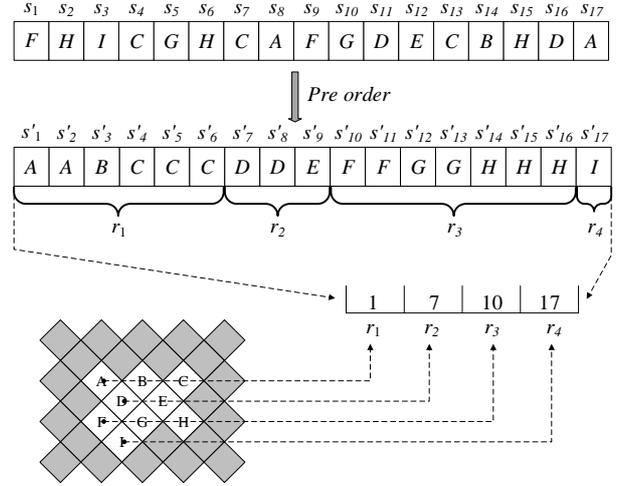}
      \caption{The construction of the two level lookup table.
               The 17 sampled points can be partitioned into 9 HEALPix pixels, and rearranged according to their pixel indices, using the latitudinal ring structure. Then a second-level lookup table can be established, as $R_{start}[1] = 1$, $R_{start}[2] = 7$, $R_{start}[3] = 10$, and $R_{start}[4] = 17$.}
         \label{preorder}
   \end{figure}
\subsection{CPU-based, multi-thread pre-ordering}
As shown in Section 1, single-level lookup table makes it possible to parallelise the gridding, thus improving the efficiency in evaluating effective data samples \citep{winkel2016cygrid}. However, since the gridding of FAST sky survey observations involves processing data with finer resolution acquired from multiple drift scans covering a large sky area, this strategy could lead to a huge lookup table for producing a final data cube with sufficiently high resolution. In this case, the single-level table could become one of the bottlenecks of gridding performance, and is not applicable for FAST surveys. In order to solve this problem, we introduce the pre-sort method, thus saving resources through reuse of the storage space for raw data, and making cross-block access to the raw data through establishing lookup-table between input samples and the HEALPix spherical tessellation possible.
 
In this work, we adopt the CPU-based multi-thread pre-ordering algorithm and a two-level lookup table scheme proposed by our previous work \citep{luo2018hygrid}, to improve the efficiency of sample points searching, with less storage required. The two-level lookup table, based on latitudinal rings, can implement flexible conversions between the ring indices and the pixel indices. Fig.~\ref{preorder} shows the pre-ordering algorithm and the process of lookup table constructions, with details shown as follows:  

\textit{I: \textbf{Pre-ordering of raw data:}} We performed pre-ordering of the raw data based on pixel indices of the HEALPix, in order to construct a two-level lookup table more conveniently. Firstly, given any partition level $N_{side}$, for any raw data $S_n \in \mathbb{S}$ with input coordinates $(\alpha_n, \delta_n)$, one can get the pixel index $P_n \in \mathbb{P}$ to which each sampling point belongs, with the help of related HEALPix APIs. Secondly, we perform key-value sorting of $P_n$ and $n$ by non-descending order, with $n$ as the subscript of data array $P_n$. Finally, we rearrange the raw data $S_n$ as $\mathbb{S}'$ according to sorted $n$, with corresponding coordinate pairs ($\alpha_n, \delta_n$).

\textit{II: \textbf{Construction of the first-level lookup table:}} As described above, $\mathbb{S}'$ and $\mathbb{P}'$ denote the sorted sampling points and the corresponding HEALPix pixels to which they belong. Considering the arrays $\mathbb{S}'$ and $\mathbb{P}'$ to be of the same size, for a given HEALPix pixel ($P_i\in\mathbb{P}'$), the corresponding starting sampled data point ($S_{start}\in\mathbb{S}'$) should meet the following criteria: $1)$ the corresponding HEALPix pixel index should equal to $P_i$; $2)$ the array index of the starting point is the smallest among all data that satisfies criteria $1)$. Based on such principles, we construct the first-level lookup table based on the mapping $P_i \mapsto S_{start}$.

\textit{III: \textbf{Construction of the second-level lookup table:}} According to the coding rule of HEALPix, the sampled data points with the same HEALPix ring index ($R_i$) should be stored in a contiguous segment of $\mathbb{S'}$. Naturally, we can obtain the minimum ring index $R_{min}$, as well as the maximum ring index $R_{max}$ for $\mathbb{P}'$. Then iteration is performed over all the rings, searching for the starting pixel indices ($P_{start}$) for each ring. Accordingly, we construct the second-level lookup table based on the mapping $R_i \mapsto P_{start}$, and symbolise it as $R_{start}[i]$.

   \begin{figure}
   \centering
   \includegraphics[width=8.5cm]{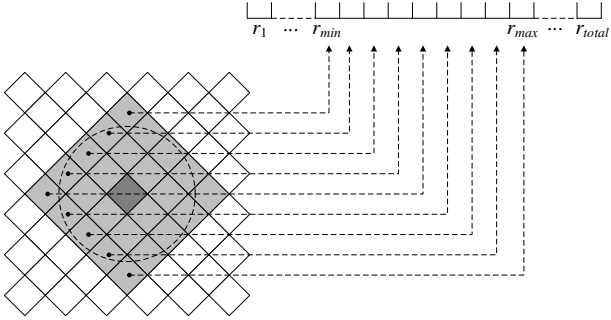}
      \caption{The searching algorithm for contributing data points. Here $r_{min}$ and $r_{max}$  are the minimum and maximum values of the latitudinal ring number falling inside the range of influence of the convolution kernel, respectively. }
         \label{search}
   \end{figure}
\subsection{Gathering-based, multi-thread GPU gridding}
The main concern of the GPU gridding process is to determine the contributing points that falling into the radius of influence by the gridding kernel. Such data are considered to be contributing to the target output grid cells. And then for each target output grid cell, the algorithm traverses the corresponding contributing input data, and performs the convolutional computations. Fig.~\ref{search} shows the searching process of the contributing points based on the two-level lookup table described in Section 3.3. The detailed steps are listed as follows:
\begin{enumerate}
      \item The minimum and maximum ring indices $R_{min}$ and $R_{max}$ within the range of influence of the convolution kernel are computed based upon the kernel size.
      \item Iteration is performed over each latitude ring $R_i \in [R_{min},R_{max}]$ to search for the corresponding starting contributing pixel in $\mathbb{P}'$ with array index falls within range of $[R_{start}[i],\ R_{start}[i+1] - 1]$.
      \item Search for starting contributing data sample $S_{start}$ within each contributing HEALPix pixel, with the help of the first-level lookup table.
      \item Traverse all the contributing data samples located within $R_i$ in $\mathbb{S}'$, starting  with $S_{start}$.
      \item Convolutional computations are performed with contributing data samples for each target grid.
\end{enumerate}

\subsection{The GPU-optimisation strategy}
We implemented GPU gridding in their native programming environment, CUDA for NVIDIA GPUs, with a runtime system and a set of C/C++ extensions. Compared with CPU, GPU usually boast a much higher amount of computing cores, in the form of Streaming Processors (SP), which can be further organised into small groups (Streaming Multiprocessors, SMs). SM is the core of the GPU architecture, the resource of the SM determined the performance of the GPU when using different thread management schemes. The optimisation of computing performance for NVIDIA GPU includes two sides: thread-based optimisation and memory-based optimisation.

\begin{figure}
   \centering
   \includegraphics[width=8.5cm]{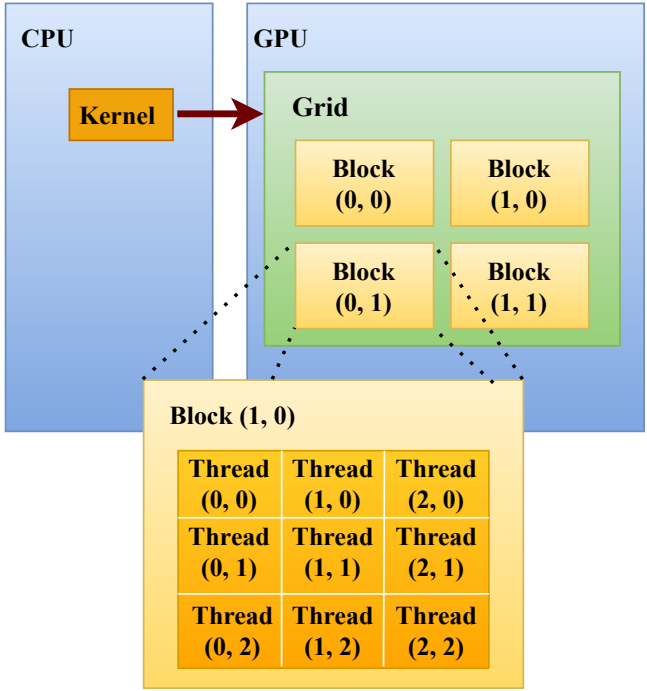}
      \caption{The basic structure of the CUDA two-level thread hierarchy. This example includes several two-dimension blocks nested in a two-dimensional grid. The dimensions of the grid and the block are determined by the built-in variables $gridDim$ and $blodkDim$, respectively.}
         \label{cuda_arc}
   \end{figure}

\textit{\textbf{I: Thread management:}}
NVIDIA proposed the concept of the hierarchical structure of threads, to facilitate thread organisation. The hierarchical structure is a two-level thread hierarchy, consisting of thread block and thread grid \citep{sanders2010cuda}. CUDA can organise three-dimensional grids and blocks, with a two-dimensional grid containing several two-dimensional blocks shown in Fig. \ref{cuda_arc} as an example \citep{cheng2014professional}. And the dimensions of the grids and blocks are determined by the built-in variables gridDim and blockDim, respectively. Each component of the built-in variables can be obtained through its x, y, and z numeric fields, that is, blockDim.x, blockDim.y, and blockDim.z. When a process is executed by threads in the thread grid level, the GPU spawns a set of grid consisting of thread blocks (both are with user-specified dimensions), and dispatches the thread blocks onto SMs \citep{veenboer2017image}. The threads in the thread blocks can be further divided as thread warps, which are the basic units for SM executions.

For any given data volume, the general steps to determine the grid and block sizes are firstly to determine the size of the blocks, and then to compute the size of the grids based on the data volume and block sizes. The sizes of the block are determined by the properties of the kernel function and the resources of GPU. Here we propose a thread organising and coarsening strategy for thread management, in order to improve the performance of multi-thread GPU gridding. The details are shown as follows:
\begin{enumerate}
    \item \textbf{Thread organisation}: We analysed the effects of thread organisation on SM executing efficiency. Since all of the input data samples located in the HEALPix rings are stored as one 1-dimensional array, the thread hierarchy we designed is consisted of one-dimensional grid and one-dimensional block, that is, gridDim.y = gridDim.z = 1, blockDim.y = blockDim.z = 1. To improve the executing efficiency of thread warps, the thread block only allocates threads on X direction, and each thread within one block is only responsible for the computing task of each target output grid cell located on the same latitude, respectively. Then the optimal number of threads in each block for the best performance is established according to various factors, including target output resolution, GPU architecture, and so on. We will make a detailed analysis of the thread organisation strategy in Section 4.2.
    
    \item \textbf{Thread coarsening}: For a higher output resolution, it is necessary to start a larger number of threads. However, due to limited resources available, not all threads can be executed on GPU simultaneously. Through analysis, we found that adjacent target output grid cells on the same latitudinal ring largely share the same set of contributing input samples. Therefore, we apply the thread coarsening technique, which utilises one thread only to calculate $\gamma$ consecutive target output pixels on the same latitudinal ring, with $\gamma$ as the thread coarsening factor. With a given factor $\gamma$, each thread only needs to perform GPU gridding once for consecutive, adjacent $\gamma$ target pixels that this thread is responsible for. In this scheme, all these target pixels share the same starting contributing input data samples, with weightings of each contributing input sample and the sum of all the weights of each target pixel stored in different registers. 
\end{enumerate}

\begin{figure}
   \centering
   \includegraphics[width=8.5cm]{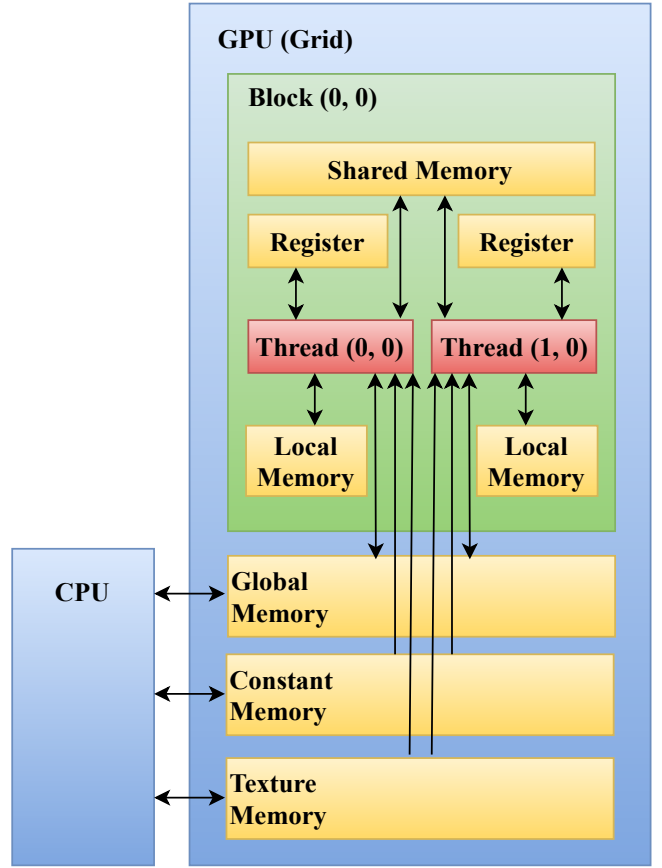}
      \caption{The hierarchy of CUDA memory. The contents in global memory can all be read and modified by every thread, while the constant and texture memories are read-only.}
         \label{cuda_ram}
   \end{figure}

\textit{\textbf{II: Memory management:}} The performance of the kernel function (here is the gridding computing on GPU) is not only depended on the execution of the thread warps, but also related to the memory access mode of CUDA. The CUDA memory model has proposed a variety of programmable memories, including registers, shared memory, local memory, and so on. Fig. \ref{cuda_ram} shows the hierarchy structure of the CUDA memory, from the picture, one can conclude that different memory mode has different scope in the thread grid. 

In our work, we utilise different types of CUDA memories for HCGrid's data storage. The optimisation strategies include:
    \begin{enumerate}
        \item The raw data, gridding results and the first-level lookup table are stored in the global memory;
        \item Texture memory is a type of global memory accessed through specially designated, read-only cache. Data in texture memory are stored globally, which means that they can be accessed by all threads. The hardware design of this type of cache here guarantees high speed access of 1-, 2-, and 3-dimensional data arrays in texture memory. Generally, such memories are suitable for image processing and storage of lookup tables. Thus, the second-level lookup table (1D) is stored in an array structure in one-dimensional texture memory, to speed up data access.
        \item Constant memory achieves the best performance with all threads in the warps reading data from the same memory address. Considering all threads in one warp use the same set of gridding parameters, e.g., kernel size, output resolution, etc., to perform the same computations with different data, we choose to store such parameters in the constant memory. 
        \item In every thread, for a given output grid cell, the weight and the kernel-weighted value (``local output'') of each contributing input data point should be stored one by one with designated arrays, only to be summed up to get the final cell reading once all contributing samples have been computed. We utilise the register to store these local output data, thus reducing the global memory access.
    \end{enumerate}
\subsection{Basic scheme of HCGrid}
In this section, we present the details of implementations of HCGrid. The whole gridding process begins with  loading raw samples, as well as designated target grid into the framework. Then the relevant parameters of the convolution kernel should be set, and the input data be pre-sorted with CPU. Finally, the GPU-based convolutional computings are performed.
\begin{enumerate}
    \item \textbf{Loading raw data:} HCGrid requires observed data and corresponding coordinates pairs as input. In our preliminary tests, we adopt the original FITS files recorded by FAST as input data format, and transition into HDF5 \citep{folk2011overview} data format is still in progress, since the FAST sky survey pipeline utilise HDF5 as intermediate format \citep{DBLP:conf/ica3pp/JiYXTWZ19}.
    \item \textbf{Reading target output map:} The most suitable size of the target grid is determined according to the area of the target sky coverage, as well as the beam width of the telescope. Those parameters are input as settings for empty target grid of HCGrid.
    \item \textbf{Initialisation of convolutional kernel:} For the specific application scenarios, the shape of the convolution kernel can lead up to different results of the gridding process, or even affect the gridding performance, depending on running environment. \cite{o1985fast} noted that the optimal convolutional kernel should be in the form of a sinc function with infinite length. However, such a function can bring huge computational loads. Thus, convolutional functions with finite lengths are adopted more often. In this work, the Gaussian function is adopted as our default kernel. The parameter initialisation of the kernel function mainly follows a previous relevant work, \cite{winkel2016cygrid}. Let $sphere\_ radius$ be the range of influence of the convolutional kernel (rather than its half-width), it is obvious that more accurate results can be obtained with a larger $sphere \_ radius$. However, in case of large amount of data, a larger range can pose great challenges to computing resources. As suggested by \cite{winkel2016cygrid}, with radially symmetric Gaussian function as the gridding kernel, usually the $sphere \_ radius$ should be set as some value between $3\sigma_{kernel} - 5\sigma_{kernel}$, depending on desired resolution, where $\sigma_{kernel}$ is the standard deviation of the kernel function. The value of $\sigma_{kernel}$ should be set according to resolution of raw data. Assuming the resolution of the telescope to be $\sigma_{data}$, as shown in \cite{winkel2016cygrid}, $\sigma_{kernel} \approx 0.5\sigma_{data}$ should be adopted to get reasonable output. In this case, the resolution of gridded data should be $\sigma_{data}^{gridded} \approx 1.12\sigma_{data}$. Here, we choose $\sigma_{data} \approx 2.'9$, which is the approximate beam size of FAST at $~1.42 GHz$. And the resolution of HEALPix grid $hpx\_max\_resolution$ is directly related to $\sigma_{kernel}$. For example, if $sphere \_ radius$ is set to be $3\sigma_{kernel}$, $hpx\_max\_resolution$ should be chosen as $\sim \sigma_{kernel}/2$. 
    \item \textbf{Pre-ordering of raw data:} HCGrid employed four interfaces for raw data pro-ordering, including the C++ STL sort, parallel\_stable\_sort and block\_indirect\_sort, both based on Boost\footnote{\url{https://www.boost.org/}}, and sort\_by\_key of Thrust\footnote{\url{https://thrust.github.io/}}. The first three are sorting interfaces for CPUs, while the last one performs GPU sortings. Through comparative analysis, it is found that block\_indirect\_sort can provide the highest capability, thus making it the default choice of pre-ordering interface.
    \item \textbf{GPU-based convolutional computation:} As mentioned in Section 3.5, we optimise the performance of GPU-based convolution computing from two aspects, thread as well as memory managements. Detailed analysis of such strategies will be analysed in Section 4.2.
\end{enumerate}

\section{Experiments and Results}\label{experiments}
In this section, we perform benchmark analysis of HCGrid. At present, since the calibration and RFI flagging \cite{yang2020deep} codes for the FAST spectral line data reduction pipeline is still in developments, it is difficult to conduct a full-scale test of HCGrid with observed data only. Thus, in this work, we mainly use simulated data to demonstrate code performance. And by comparing with the accurate results of Cygrid to illustrate the correctness of the HCGrid's gridding results. We generate simulated data in the similar way as \cite{winkel2016cygrid}, that is, one Gaussian-distributed random value is assigned to each evenly-distributed random position $n$ with coordinates $(\alpha_n, \delta_n)$ in a pre-defined field of view.
\subsection{Experimental setings}	
As listed in Table~\ref{table:environment}, we carried out our experiments on three workstations with different GPU hardware architectures. These workstations can be referred to as K40 (hosting an NVIDIA Tesla K40 GPU), T4 (equipped with an NVIDIA Tesla T4 GPU), and V100 (with an NVIDIA Tesla V100 GPU). We have adopted CUDA 8.0.61, GPU driver version 390.116 for K40; CUDA 8.0.61, GPU driver version 440.33.01 for T4; and CUDA 8.0.61 with GPU driver version 440.33.01 for V100.

To achieve performance optimisation, we made a full consideration of the resource allocation under different GPU architecture, thus reasonably allocate resources to achieve relatively better performance.
   
Details of our \textbf{HCGrid} performance analysis include:
   \begin{enumerate}
       \item We analyse the effects of thread organisation strategies on gridding performance, with sampling space of input data, output resolution, and the amount of the input data fixed.
       \item We analyse the effects of the thread coarsening factors on gridding performance at different output resolutions, with the sampling space coverage and the input data amount fixed.
       \item We analyse the dependency of HCGrid computing time on input data amounts, with sampling space coverage and output resolution remaining fixed.
       \item We analyse the dependency of computing time on target grid coverage, with the input data amount and output resolution of output fixed.
       \item We make a relative comparison between the Cygrid and HCGrid to illustrate the advantage of the GPU on the gridding process.
   \end{enumerate}
\begin{table*}
\centering    
\renewcommand\arraystretch{1.5}
\caption{The configurations of the workstation for experiments} 
\label{table:environment}
\begin{tabular}{cccccccccc}
\hline\hline
    Model  & Type        & Architecture & \begin{tabular}[c]{@{}c@{}}Clock\\ (GHz)\end{tabular} & Cores & SMs & \begin{tabular}[c]{@{}c@{}}Mem size\\ (GB)\end{tabular} & \begin{tabular}[c]{@{}c@{}}Mem bw\\ (GB/s)\end{tabular} & \begin{tabular}[c]{@{}c@{}}CUDA\\ Capability\end{tabular} \\ \hline
    Intel Xeon E5-2620   & CPU    & Sandy-EP   & 2.4  & 6     & -    & 32  & 68.27    & - \\                                                       
     NVIDIA Tesla K40     & GPU    & Kepler   & 0.745   & 2880  & 15   & 12  & 288.4  & 3.5 \\ 
     \hline
    Intel Xeon Platinum 8255C & CPU    & Cascade Lake   & 2.5  & 8     & -    & 32  & 140.8    & - \\                                                       
     NVIDIA Tesla T4           & GPU    & Turing   & 0.585   & 2560  & 40   & 16  & 320  & 7.5 \\ 
     \hline
     Intel Xeon Silver 4114   & CPU    & Skylake-EP   & 2.2  & 16     & -    & 32  & 115.21    & - \\                                                       
     NVIDIA Tesla V100     & GPU    & Volta   & 1.246   & 5120  & 80   & 16  & 897.0  & 7.0 \\ 
     \hline
     
\end{tabular}
\end{table*}

\subsection{Performance analysis} 

   \begin{figure}
   \centering
   \includegraphics[width=8cm]{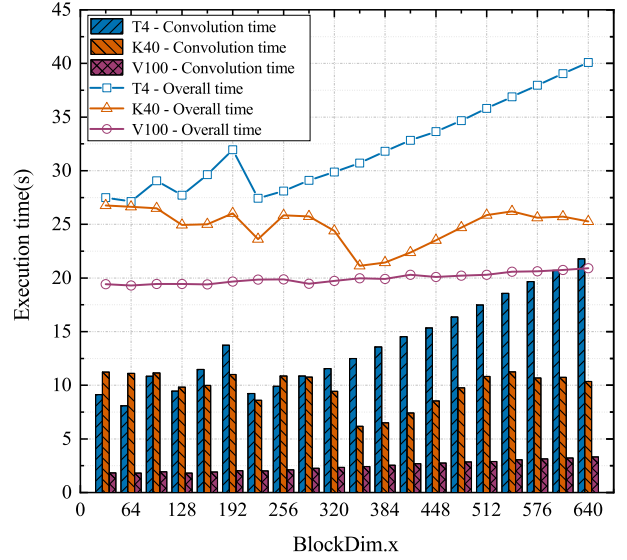}
      \caption{Benchmark results of thread organisation strategy.
                 With $10^8$ input samples, and the thread coarsening factor $\gamma = 1$, the processing time of GPU convolution, as well as HCGrid processing time changes with different settings of thread organisation strategy. 
              }
         \label{Block}
   \end{figure}

	\textit{\textbf{1) Performance vs thread organisation strategy: }}
	With a field size of $ 5^\circ \times 5^\circ $, the output resolutions of $\sigma_{grid}=200''$(which is approximately the average beam size of the FAST across the observing band of its 19-beam receive), and input sample amount of $N=10^8$, we implement various thread configurations with different GPU architectures by configuring the values of thread grids and thread blocks, to analyse the impact of the choice of thread organisation strategies on the execution time of GPU convolution, as well as the overall execution time of HCGrid. 
	
	Take workstation K40 with a CUDA capability of 3.5 as an example. Here the total number of registers available for each block of is $64K$, while our compilation report shows that the core function of \textbf{HCGrid} utilises 184 registers, without using shared memory to store parameters. Thus, it is expected that each thread block can execute $\sim 64K/184 \approx 356$ threads simultaneously. Fig.~\ref{Block} shows that the execution time varies with thread organisation methods. With the number of blocks along X-dimension $blockdim.x = 352$, and the number of grids in the same dimension ($griddim.x$) as 25, the fastest execution speed can be obtained. That is because, in this case, the value of $blockdim.x$ is close to the maximum thread count 356, thus enabling the full utilisation of the computing capability of the K40 GPU.
	
	And for workstation T4 with a CUDA capability of 7.5, the total number of registers available for each block is also $64K$.	However, the number of SPs in T4 is 2560, which is smaller than K40's 2880. That is, T4 can execute 2560 threads at most at the same time. Partially due to this reason, the execution time of HCGrid on T4 is longer than that of K40. And since each SM of T4 has 64 SPs, one SM of T4 can only execute 64 threads simultaneously. Thus, if we allocate $blockdim.x = 352$ as we did to K40, the thread waiting should be expected, and thread scheduling overhead could be increased to some extent. Therefore, experiments show that the best computing performance can be obtained for $blockdim.x=64$. In this configuration, it is possible to achieve maximum thread parallelisation for each SM with minimal thread scheduling overhead. 
	
	Similarly, for workstation V100 with CUDA capability of 7.0, and the largest SP among all our workstations, a prominent advantage in terms of thread organisation, with the shortest GPU convolution time can be achieved. And although the architecture of V100 is different from T4, the number of SPs in each SM of these two GPUs are the same. Thus, the thread organisation strategies for T4 can be adopted for V100, with $blockdim.x$ remains to be 64 for the best performance. It can be seen in Fig.~\ref{Block} that our conclusions have been confirmed.
    
    Based on the analysis above, for the thread organisation configuration, the architecture of the GPU and the number of SPs in the SM should be carefully adjusted, in order to select the most appropriate scheme to improve the performance of GPU parallelisation. For mainstream GPU architectures by NVIDIA, including Turing, Volta, Pascal, Kepler, Fermi, and Maxwell, the minimum number of SPs in each SM equals to 32 (for Fermi architecture). Thus, when taking thread configuration into consideration only, we get the empirical Eq.~(\ref{thread_equation}) as follows:
   \begin{equation}
      T_{max} = (Register\_num)/184 
    \label{Register}
   \end{equation}
   
    \begin{eqnarray}
    blockdim.x = 
    \begin{cases}
    SP                      & 32 \leq SP < \frac{1}{2}T_{max} \\
    T_{max}                 & SP \geq \frac{1}{2}T_{max} \\
    other                   & Based\ on\ actual\ test\ results 
    \end{cases}
    \label{thread_equation}
    \end{eqnarray}
    In Eq.~(\ref{Register}), $Register\_num$ represents the total number of registers available for each thread block of the GPU, and $T_{max}$ is the maximum number of threads that each thread block can execute simultaneously when running HCGrid. In Eq.~(\ref{thread_equation}), $ SP $ is the number of SPs in each SM of the GPU. For example, for K40, $T_{max} = 356$, and the number of SPs in each SM is 192, which is greater than $\frac{1}{2}T_{max}$. Thus, the value of $blockdim.x$ should be set as 352. However, since thread configuration may also be affected by various factors, such as GPU clock frequency and memory bandwidth, the GPU thread organisation parameters should be adjusted according to the specific computing environment to get optimised results.
    
\begin{figure}
   \centering
   \includegraphics[width=\hsize]{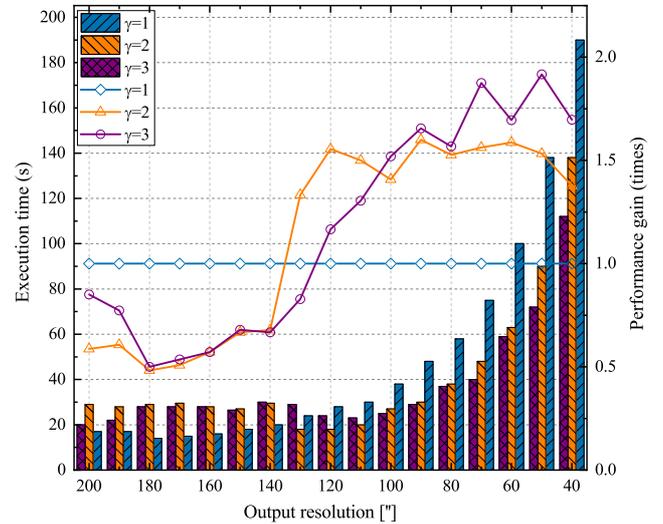}
      \caption{The processing time and performance gain of the convolution part on GPU. The histogram represents the execution time, and the curves show the performance gain. In our calculations, $10^8$ samples are gridded onto a target field of $ 5^\circ \times 5^\circ $, with different thread coarsening factor $\gamma$.
               }
         \label{thread_coarsening}
   \end{figure}

	 \textit{\textbf{2) Performance vs thread coarsening strategy: }} when make thread coarsening experiment on K40 workstations, Fig.~\ref{thread_coarsening} shows the execution time for convolution with different thread coarsening factors ($\gamma$ = 1,2,3) at different output grid resolutions. The relative performance improvements for each run relative to the processing time for the $\gamma$ = 1 case are also shown in the same figure. It can be seen that the finner the output grid resolution, the more improvement of gridding performance can be obtained with thread coarsening.

    When starting a large number of threads, on the one hand, it is possible that not all threads can be run concurrently on the GPU, due to limited resources available. On the other hand, in this case, the workload of thread scheduling would be increased, resulting in performance degradations. Thus, thread coarsening technique is often applied, with multiple input samples handled by one thread only. As shown in Fig.~\ref{thread_coarsening}, With $\gamma = 2$ or 3, the number of starting threads is reduced, hence the scheduling workload can be reduced, which can lead to better gridding performance as a result.
    
    We have also performed thread coarsening experiments on T4 and V100 workstations, with similar results to K40 obtained. Suggestions for thread coarsening parameter settings are listed as follows: 
    \begin{enumerate}
        \item With $ 150''\leq\sigma_{data}\leq200''$, it is reasonable to set $\gamma=1$ for better computing performance;
        \item If $ 110''\leq\sigma_{data}\leq140''$,  a more reasonable choice should be $\gamma=2$;
        \item $\gamma=3$ works better for $ 40''\leq\sigma_{data}\leq100''$.
        
    \end{enumerate}

\begin{figure}
   \centering
   \includegraphics[width=7.5cm]{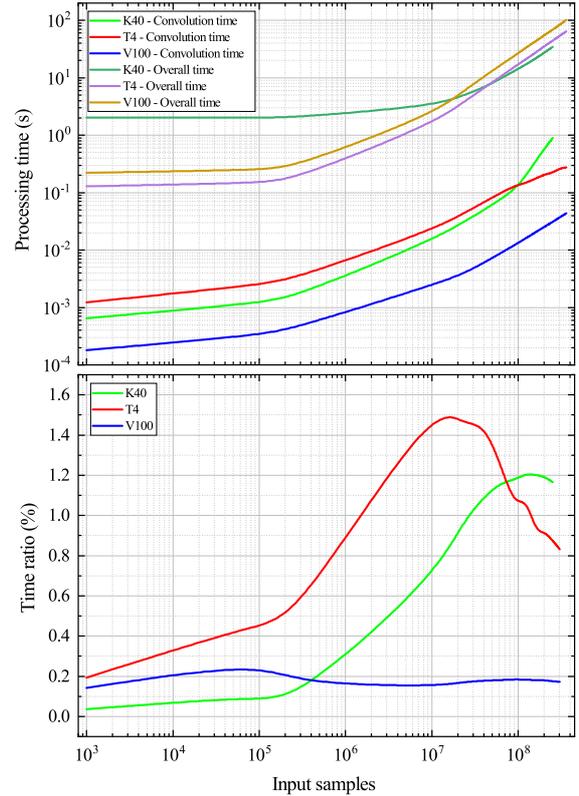}
      \caption{Results of computational complexity analysis. The top panel shows the HCGrid processing time for different input data volume, while the lower one presents the ratio of time for convolutional computations to the whole executing time.}
         \label{sample_processing_time}
   \end{figure}

	\textit{\textbf{3) Performance vs input data volume:}} With a input data sky coverage of $ 5^{\circ} \times 5^{\circ} $, and a target resolution of $\sigma_{grid}=200''$, we have $ I \times J = \left(5^{\circ}/200''\right)\times \left(5^{\circ}/200''\right) = 90 \times 90$. The processing time of GPU convolution and the overall computing time of HCGrid on three workstations with different configurations for input data with $10^3 \sim 3.6 \times 10^8$ samples are shown in the top panel of Fig.~\ref{sample_processing_time}. It can be seen that the processing time of HCGrid increases quasi-linearly with the input data size. If the number of input samples is with an order of magnitude less than $10^5$, the HCGrid computational complexity is $O\left(1\right)$; while with samples larger than $10^5$, the complexity changes to $O\left(n\right)$. Thus, it can be concluded that the computational complexity of HCGrid should lie somewhere between $O\left(1\right)$ and $O\left(n\right)$. And the lower panel of Fig.~\ref{sample_processing_time} presents the ratio of the convolution time to the whole processing time of HCGrid, with a highest reading as $1.5\%$ for the workstation with T4 GPU. In one word, with the help of GPU, it is demonstrated that the convolutional computations, which should be the most computationally intensive task in the gridding process, is no longer a time-consuming component for HCGrid.

\begin{figure}
   \centering
   \includegraphics[width=7.5cm]{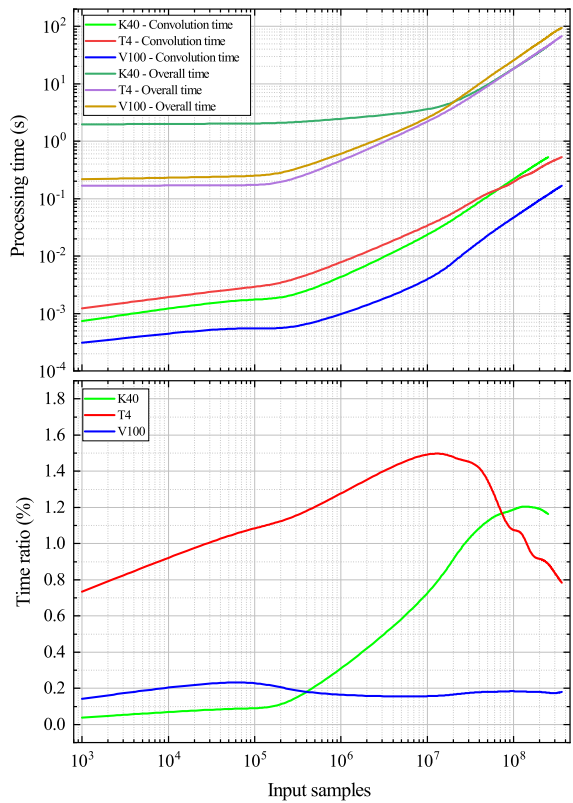}
      \caption{Analysis of HCGrid performance in relationship with field size. The filed size varies from $0.1^\circ \times 0.1^\circ$ to $60^\circ\times 60^\circ$, with input sample density as $10^5$ per square degree.
              }
         \label{filed_processing_time}
   \end{figure}

    \textit{\textbf{4) Performance vs filed size:}} To further analyse the impact of different field sizes, we conduct similar experiments, as shown in  Fig.~\ref{sample_processing_time}. Here the size of the field is varies from $0.1^\circ \times 0.1^\circ$ to $60^\circ \times 60^\circ$, with the input data amount as $10^5$ samples per square degree. As depicted in Fig.~\ref{filed_processing_time}, the test results are similar to Fig.~\ref{sample_processing_time}, which demonstrated that the processing time of HCGrid is mostly influenced by input data volume, with limited relevance to the sizes of the field.

\subsection{Comparison with Cygrid}
    As mentioned above, Cygrid is a high-performance gridding implementation for CPUs. By comparison, the GPU-implemented HCGrid proposed by this work improves the hardware utilisation, adding with the introducing of the pre-sorting operation of the input samples. This section gives a relative comparison of HCGrid and Cygrid performance, in order to demonstrate the advantage of GPU for gridding operations.
    
\begin{figure}
   \centering
   \includegraphics[width=7.5cm]{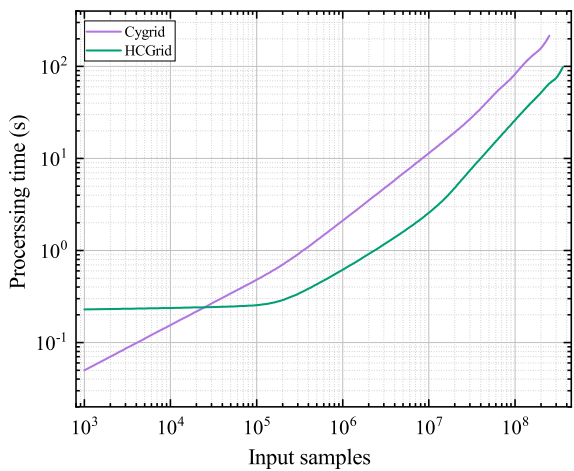}
      \caption{Comparison of Cygrid and HCGrid. The input data volume ranged from $10^3 \sim 3.6 \times 10^8$ sample. Here the executing platform for Cygrid is 16 CPU cores from the V100 workstation, while HCGrid is deployed on the same workstation, utilising its GPU.}
         \label{comopare}
   \end{figure}

    The grid coverage on the target field for this experiment is $5^\circ \times 5^\circ$, with a grid resolution of $\sigma_{grid} = 200''$, and the gridding kernel width of $\vartheta_{fwhm} = 300''$. The input data volume varies from $10^3 \sim 3.6 \times 10^8$ simulated samples. The Cygrid is executed on the V100 workstation with 16 processor cores, while HCGrid is deployed on the same computing environment with GPU exploited. Fig.~\ref{comopare} compared the executing time of Cygrid and HCGrid, it can be seen that
    the computational complexity of Cygrid, which is closer to $O\left(n\right)$ \citep{winkel2016cygrid}, has been verified by the trend in the corresponding curve. And with the number of input samples exceeding  $10^5$, the executing speed of HCGrid is several times faster than Cygrid. However, when dealing with smaller batches of input, data transmissions between CPU and GPU cause notable degradation in performance of HCGrid, resulting in a slower speed than Cygrid. In case of larger amount of input data, the performance improvements due to GPU-implemented convolutions far exceeds the overhead of such data transmissions, leading to significantly better overall performance of HCGrid.

\begin{figure*}
   \centering
   \includegraphics[width=\hsize]{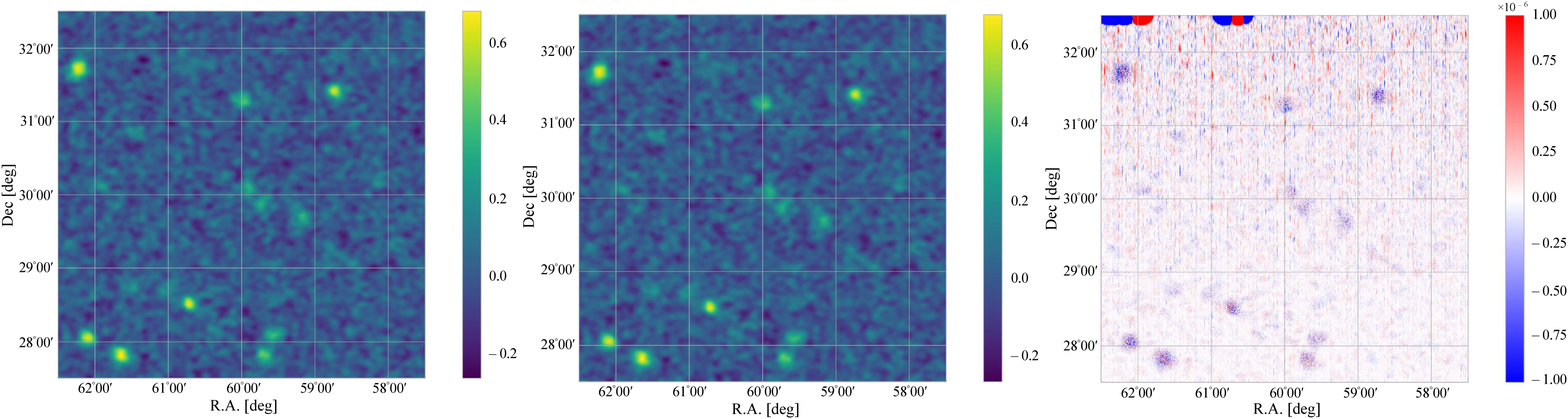}
      \caption{The gridding results of Cygrid (left) and HCGrid (middle), and the differential map between two sets of results (right) for a $5^\circ \times 5^\circ $ sampling space with 25 point sources.}
         \label{Source25}
   \end{figure*}

\begin{figure*}
   \centering
   \includegraphics[width=\hsize]{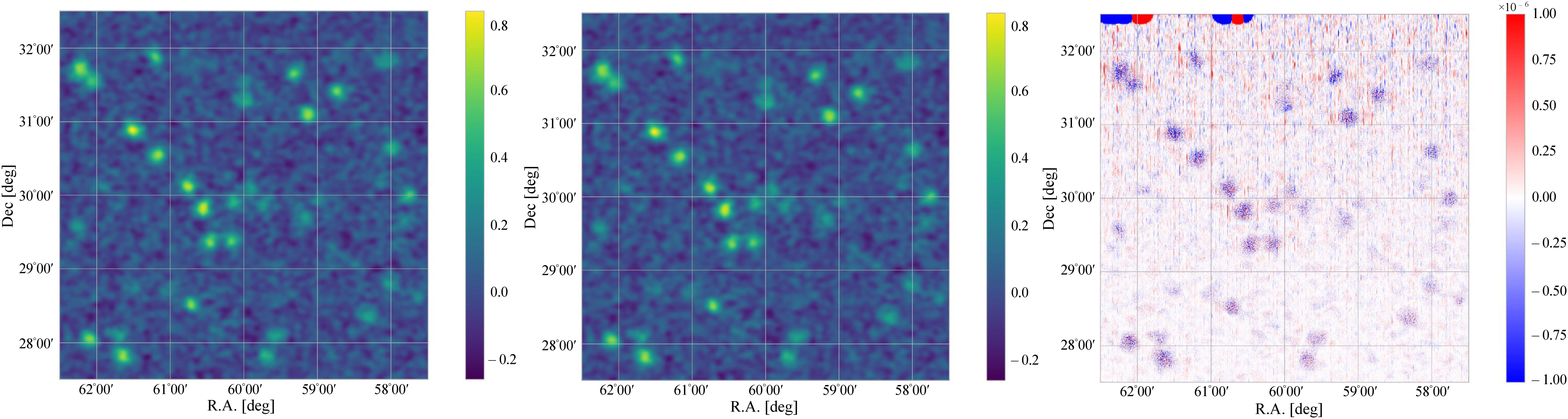}
      \caption{The gridding results of Cygrid (left) and HCGrid (middle), and the differential map between two sets of results (right) for a $5^\circ \times 5^\circ $ sampling space with 50 point sources.}
         \label{Source50}
   \end{figure*}

    Meanwhile, to verify the correctness of gridding results, we adopt the results produced by Cygrid, which is a mature and widely used gridding module, as our benchmark. For a $5^\circ \times 5^\circ$ sampling space with $10^6$ data samples and the same set of parameters (including beam size, convolution kernel, etc.), Figs.~\ref{Source25} and \ref{Source50} show the gridding products and the corresponding differential maps between HCGrid and Cygrid for 25 and 50 point sources, respectively. It can be seen that the results produced by HCGrid are generally compatible with those from Cygrid, with all sources clearly reconstructed and resolved. The differential maps show that the difference between two sets of results, which can be explained by different hardware architectures between CPU and GPU, can be largely neglected.

\section{Conclusions}
    In this paper, we introduce a convolution-based gridding framework \textbf{HCGrid} for radio astronomy in hybrid computing environments, in order to meet the requirements of large single-dish radio telescopes such as FAST. \textbf{HCGrid} features the first implementation of convolution-based gridding with gather strategy for CPU-GPU heterogeneous platform.

    To increase the searching efficiency of raw input samples, \textbf{HCGrid} makes full use of advantages of both CPU and GPU. It implemented a fast parallel ordering algorithm based on HEALPix on CPU, established a two-level lookup table to speed up the sample searching process, and accelerated the convolution operation using GPU with several optimisation strategies.
    
    To ensure the scalability of HCGrid, we have performed experiments on three workstations with various GPU architectures. All of the performance optimisation strategies have been tested on all three architectures, with general guidance for related performance parameter optimisations presented. And compared with CPU implementations of gridding process, HCGrid has the performance advantage due to GPU-based carried out the time-consuming convolution computing on the GPU.
    
    Further improvements of the \textbf{HCGrid} is expected, as minor bugs being further tested and removed, and related user manual being continuously updated. Also, the time-consuming data transmissions between CPU and GPU, as noted in Section 4.3, is under further investigations with various memthods, including the CUDA streaming technology, which will improve the transmission efficiency. The development of multi-channel spectral data handling capabilities, as well as the corresponding I/O  optimisation works are still underway. And the final integration of HCGrid to the data reduction pipeline of FAST telescope will also be implemented in the near future.

\section*{Acknowledgements}

We thank the Cygrid team (B. Winkel, L.Fl\"{o}er, D. Lenz) for providing the Cython code of \textbf{Cygrid}. Also, since the HCGrid code is developed ultilising the HEALPix library \citep{gorski2005healpix}, and the Python/C++ open source libraries including NumPy (Walt et al. 2011), SciPy (Jones et al. 2001), Astropy (Astropy Collaboration 2013), as well as matplotlib (Hunter 2007), we would like to thank all the developers for these packages.
This work is supported by the Joint Research Fund in Astronomy (grant Nos. U1731125, U1731243) under a cooperative agreement between the National Natural Science Foundation of China (NSFC) and Chinese Academy of Sciences, NSFC grant No. 11903056, the Cultivation Project for FAST Scientific Payoff and Research Achievement of CAMS-CAS, as well as the Open Project Program of the Key Laboratory of FAST, NAOC, Chinese Academy of Sciences.

\section*{Data Availability}

The test data adopted for benchmarking HCGrid are generated by the authors using algorithms described in Section \ref{experiments}, in reference to \cite{winkel2016cygrid}. The test data can be accessed by sending request to the authors.




\bibliographystyle{mnras}
\bibliography{example} 

\begin{thebibliography}{}
\makeatletter
\relax
\def\mn@urlcharsother{\let\do\@makeother \do\$\do\&\do\#\do\^\do\_\do\%\do\~}
\def\mn@doi{\begingroup\mn@urlcharsother \@ifnextchar [ {\mn@doi@}
  {\mn@doi@[]}}
\def\mn@doi@[#1]#2{\def\@tempa{#1}\ifx\@tempa\@empty \href
  {http://dx.doi.org/#2} {doi:#2}\else \href {http://dx.doi.org/#2} {#1}\fi
  \endgroup}
\def\mn@eprint#1#2{\mn@eprint@#1:#2::\@nil}
\def\mn@eprint@arXiv#1{\href {http://arxiv.org/abs/#1} {{\tt arXiv:#1}}}
\def\mn@eprint@dblp#1{\href {http://dblp.uni-trier.de/rec/bibtex/#1.xml}
  {dblp:#1}}
\def\mn@eprint@#1:#2:#3:#4\@nil{\def\@tempa {#1}\def\@tempb {#2}\def\@tempc
  {#3}\ifx \@tempc \@empty \let \@tempc \@tempb \let \@tempb \@tempa \fi \ifx
  \@tempb \@empty \def\@tempb {arXiv}\fi \@ifundefined
  {mn@eprint@\@tempb}{\@tempb:\@tempc}{\expandafter \expandafter \csname
  mn@eprint@\@tempb\endcsname \expandafter{\@tempc}}}

\bibitem[\protect\citeauthoryear{{Calabretta} \& {Greisen}}{{Calabretta} \&
  {Greisen}}{2002}]{calabretta2002representations}
{Calabretta} M.~R.,  {Greisen} E.~W.,  2002, \mn@doi [\aap]
  {10.1051/0004-6361:20021327}, \href
  {https://ui.adsabs.harvard.edu/abs/2002A&A...395.1077C} {395, 1077}

\bibitem[\protect\citeauthoryear{{Cheng}, {Grossman}  \& {McKercher}}{{Cheng}
  et~al.}{2014}]{cheng2014professional}
{Cheng} J.,  {Grossman} M.,   {McKercher} T.,  2014, Professional CUDA C
  Programming.
John Wiley \& Sons - New York

\bibitem[\protect\citeauthoryear{{Folk}, {Heber}, {Koziol}, {Pourmal}  \&
  {Robinson}}{{Folk} et~al.}{2011}]{folk2011overview}
{Folk} M.,  {Heber} G.,  {Koziol} Q.,  {Pourmal} E.,   {Robinson} D.,  2011, in
  Proceedings of the EDBT/ICDT 2011 Workshop on Array Databases. pp 36--47,
  \mn@doi{10.1145/1966895.1966900}

\bibitem[\protect\citeauthoryear{{Giovanelli} et~al.,}{{Giovanelli}
  et~al.}{2005}]{giovanelli2005arecibo}
{Giovanelli} R.,  et~al., 2005, \mn@doi [\aj] {10.1086/497431}, \href
  {https://ui.adsabs.harvard.edu/abs/2005AJ....130.2598G} {130, 2598}

\bibitem[\protect\citeauthoryear{{G{\'o}rski}, {Hivon}, {Banday}, {Wand elt},
  {Hansen}, {Reinecke}  \& {Bartelmann}}{{G{\'o}rski}
  et~al.}{2005}]{gorski2005healpix}
{G{\'o}rski} K.~M.,  {Hivon} E.,  {Banday} A.~J.,  {Wand elt} B.~D.,  {Hansen}
  F.~K.,  {Reinecke} M.,   {Bartelmann} M.,  2005, \mn@doi [\apj]
  {10.1086/427976}, \href
  {https://ui.adsabs.harvard.edu/abs/2005ApJ...622..759G} {622, 759}

\bibitem[\protect\citeauthoryear{Ji, Yu, Xiao, Tang, Wang  \& Zhang}{Ji
  et~al.}{2019}]{DBLP:conf/ica3pp/JiYXTWZ19}
Ji Y.,  Yu C.,  Xiao J.,  Tang S.,  Wang H.,   Zhang B.,  2019, in Wen S.,
  Zomaya A.~Y.,   Yang L.~T.,  eds,  Lecture Notes in Computer Science Vol.
  11945, Algorithms and Architectures for Parallel Processing - 19th
  International Conference, {ICA3PP} 2019, Melbourne, VIC, Australia, December
  9-11, 2019, Proceedings, Part {II}. Springer, pp 656--672,
  \mn@doi{10.1007/978-3-030-38961-1\_55}

\bibitem[\protect\citeauthoryear{{Kalberla}, {Burton}, {Hartmann}, {Arnal},
  {Bajaja}, {Morras}  \& {P{\"o}ppel}}{{Kalberla}
  et~al.}{2005}]{kalberla2005leiden}
{Kalberla} P.~M.~W.,  {Burton} W.~B.,  {Hartmann} D.,  {Arnal} E.~M.,  {Bajaja}
  E.,  {Morras} R.,   {P{\"o}ppel} W.~G.~L.,  2005, \mn@doi [\aap]
  {10.1051/0004-6361:20041864}, \href
  {https://ui.adsabs.harvard.edu/abs/2005A&A...440..775K} {440, 775}

\bibitem[\protect\citeauthoryear{{Kalberla} et~al.,}{{Kalberla}
  et~al.}{2010}]{kalberla2010gass}
{Kalberla} P.~M.~W.,  et~al., 2010, \mn@doi [\aap]
  {10.1051/0004-6361/200913979}, \href
  {https://ui.adsabs.harvard.edu/abs/2010A&A...521A..17K} {521, A17}

\bibitem[\protect\citeauthoryear{{L{\'e}na}, {Rouan}, {Lebrun}, {Mignard}  \&
  {Pelat}}{{L{\'e}na} et~al.}{2012}]{lena2012observational}
{L{\'e}na} P.,  {Rouan} D.,  {Lebrun} F.,  {Mignard} F.,   {Pelat} D.,  2012,
  Observational Astrophysics.
Springer - Verlag Berlin Heidelber

\bibitem[\protect\citeauthoryear{{Li} et~al.,}{{Li} et~al.}{2018}]{Li2018}
{Li} D.,  et~al., 2018, \mn@doi [IEEE Microwave Magazine]
  {10.1109/MMM.2018.2802178}, \href
  {https://ui.adsabs.harvard.edu/abs/2018IMMag..19..112L} {19, 112}

\bibitem[\protect\citeauthoryear{Luo, Xiao, Yu, Bi, Ji, Sun, Zhang  \&
  Wang}{Luo et~al.}{2018}]{luo2018hygrid}
Luo Q.,  Xiao J.,  Yu C.,  Bi C.,  Ji Y.,  Sun J.,  Zhang B.,   Wang H.,  2018,
  in International Conference on Algorithms and Architectures for Parallel
  Processing. pp 621--635, \mn@doi{10.1007/978-3-030-05051-1_43}

\bibitem[\protect\citeauthoryear{{McCool}, {Reinders}  \& {Robison}}{{McCool}
  et~al.}{2012}]{mccool2012structured}
{McCool} M.,  {Reinders} J.,   {Robison} A.,  2012, Structured parallel
  programming: patterns for efficient computation.
Elsevier

\bibitem[\protect\citeauthoryear{{Merry}}{{Merry}}{2016}]{merry2016faster}
{Merry} B.,  2016, \mn@doi [Astronomy and Computing]
  {10.1016/j.ascom.2016.05.004}, \href
  {https://ui.adsabs.harvard.edu/abs/2016A&C....16..140M} {16, 140}

\bibitem[\protect\citeauthoryear{{Mink}}{{Mink}}{2006}]{mink2006wcstools}
{Mink} D.,  2006, in {Gabriel} C.,  {Arviset} C.,  {Ponz} D.,   {Enrique} S.,
  eds,  Astronomical Society of the Pacific Conference Series Vol. 351,
  Astronomical Data Analysis Software and Systems XV. p.~204

\bibitem[\protect\citeauthoryear{{Nan}}{{Nan}}{2006}]{nan2006five}
{Nan} R.,  2006, \mn@doi [Science in China: Physics, Mechanics and Astronomy]
  {10.1007/s11433-006-0129-9}, \href
  {https://ui.adsabs.harvard.edu/abs/2006ScChG..49..129N} {49, 129}

\bibitem[\protect\citeauthoryear{{Nan} et~al.,}{{Nan}
  et~al.}{2011}]{nan2011five}
{Nan} R.,  et~al., 2011, \mn@doi [International Journal of Modern Physics D]
  {10.1142/S0218271811019335}, \href
  {https://ui.adsabs.harvard.edu/abs/2011IJMPD..20..989N} {20, 989}

\bibitem[\protect\citeauthoryear{{O'Sullivan}}{{O'Sullivan}}{1985}]{o1985fast}
{O'Sullivan} J.~D.,  1985, \mn@doi [IEEE transactions on medical imaging]
  {10.1109/TMI.1985.4307723}, 4, 200

\bibitem[\protect\citeauthoryear{{Romein}}{{Romein}}{2012}]{romein2012efficient}
{Romein} J.~W.,  2012, in Proceedings of the 26th ACM international conference
  on Supercomputing. pp 321--330, \mn@doi{10.1145/2304576.2304620}

\bibitem[\protect\citeauthoryear{{Sanders} \& {Kandrot}}{{Sanders} \&
  {Kandrot}}{2010}]{sanders2010cuda}
{Sanders} J.,  {Kandrot} E.,  2010, CUDA by example: an introduction to
  general-purpose GPU programming.
Addison-Wesley - Boston

\bibitem[\protect\citeauthoryear{{Schweizer}, {Besta}  \&
  {Hoefler}}{{Schweizer} et~al.}{2015}]{schweizer2015evaluating}
{Schweizer} H.,  {Besta} M.,   {Hoefler} T.,  2015, in 2015 International
  Conference on Parallel Architecture and Compilation (PACT). pp 445--456,
  \mn@doi{10.1109/PACT.2015.24}

\bibitem[\protect\citeauthoryear{Smith, Dunning, Smart, Shaw, Mackay, Bowen  \&
  Hayman}{Smith et~al.}{2017}]{smith2017performance}
Smith S.~L.,  Dunning A.,  Smart K.~W.,  Shaw R.,  Mackay S.,  Bowen M.,
  Hayman D.,  2017, in 2017 IEEE International Symposium on Antennas and
  Propagation \& USNC/URSI National Radio Science Meeting. pp 2137--2138,
  \mn@doi{10.1109/APUSNCURSINRSM.2017.8073111}

\bibitem[\protect\citeauthoryear{{Veenboer}, {Petschow}  \&
  {Romein}}{{Veenboer} et~al.}{2017}]{veenboer2017image}
{Veenboer} B.,  {Petschow} M.,   {Romein} J.~W.,  2017, in 2017 IEEE
  International Parallel and Distributed Processing Symposium (IPDPS). pp
  545--554, \mn@doi{10.1109/IPDPS.2017.68}

\bibitem[\protect\citeauthoryear{{Wells} \& {Greisen}}{{Wells} \&
  {Greisen}}{1979}]{wells1979fits}
{Wells} D.~C.,  {Greisen} E.~W.,  1979, in {Sedmak} G.,  {Capaccioli} M.,
  {Allen} R.~J.,  eds, Image Processing in Astronomy. p.~445

\bibitem[\protect\citeauthoryear{{Winkel}, {Kerp}, {Fl{\"o}er}, {Kalberla},
  {Ben Bekhti}, {Keller}  \& {Lenz}}{{Winkel}
  et~al.}{2016a}]{winkel2016effelsberg}
{Winkel} B.,  {Kerp} J.,  {Fl{\"o}er} L.,  {Kalberla} P.~M.~W.,  {Ben Bekhti}
  N.,  {Keller} R.,   {Lenz} D.,  2016a, \mn@doi [\aap]
  {10.1051/0004-6361/201527007}, \href
  {https://ui.adsabs.harvard.edu/abs/2016A&A...585A..41W} {585, A41}

\bibitem[\protect\citeauthoryear{{Winkel}, {Lenz}  \& {Fl{\"o}er}}{{Winkel}
  et~al.}{2016b}]{winkel2016cygrid}
{Winkel} B.,  {Lenz} D.,   {Fl{\"o}er} L.,  2016b, \mn@doi [\aap]
  {10.1051/0004-6361/201628475}, \href
  {https://ui.adsabs.harvard.edu/abs/2016A&A...591A..12W} {591, A12}

\bibitem[\protect\citeauthoryear{{Yang}, {Yu}, {Xiao}  \& {Zhang}}{{Yang}
  et~al.}{2020}]{yang2020deep}
{Yang} Z.,  {Yu} C.,  {Xiao} J.,   {Zhang} B.,  2020, \mn@doi [\mnras]
  {10.1093/mnras/stz3521}, \href
  {https://ui.adsabs.harvard.edu/abs/2020MNRAS.492.1421Y} {492, 1421}

\bibitem[\protect\citeauthoryear{{van Amesfoort}, {Varbanescu}, {Sips}  \& {van
  Nieuwpoort}}{{van Amesfoort} et~al.}{2009}]{van2009evaluating}
{van Amesfoort} A.~S.,  {Varbanescu} A.~L.,  {Sips} H.~J.,   {van Nieuwpoort}
  R.~V.,  2009, in Proceedings of the 6th ACM conference on Computing
  frontiers. pp 207--216, \mn@doi{10.1145/1531743.1531777}

\makeatother
\end{thebibliography}





\bsp	
\label{lastpage}
\end{document}